\documentclass[journal]{IEEEtran}
\usepackage{cite}

\ifCLASSINFOpdf
  \usepackage[pdftex]{graphicx}
\else
  \usepackage[dvips]{graphicx}
\fi
\usepackage{amsmath}
\usepackage{algorithmic}

\usepackage{algorithm}

\usepackage{array}

\ifCLASSOPTIONcompsoc
 \usepackage[caption=false,font=normalsize,labelfont=sf,textfont=sf]{subfig}
\else
 \usepackage[caption=false,font=footnotesize]{subfig}
\fi
\usepackage{dblfloatfix}
\usepackage{url}

\usepackage{booktabs}  
\usepackage{threeparttable}  
\usepackage{multirow}  
\usepackage{xcolor} 

\begin{document}
%
\title{Robust Feature Learning on Long-Duration Sounds for Acoustic Scene Classification}
%
%
%


\author{Yuzhong~Wu and 
        Tan~Lee}

%
%

\markboth{}%
{}
%



\maketitle

\begin{abstract}

Acoustic scene classification (ASC) aims to identify the type of scene (environment) in which a given audio signal is recorded. The log-mel feature and convolutional neural network (CNN) have recently become the most popular time-frequency (TF) feature representation and classifier in ASC. An audio signal recorded in a scene may include various sounds overlapping in time and frequency. The previous study suggests that separately considering the long-duration sounds and short-duration sounds in CNN may improve ASC accuracy. This study addresses the problem of the generalization ability of acoustic scene classifiers. In practice, acoustic scene signals' characteristics may be affected by various factors, such as the choice of recording devices and the change of recording locations. When an established ASC system predicts scene classes on audios recorded in unseen scenarios, its accuracy may drop significantly. The long-duration sounds not only contain domain-independent acoustic scene information, but also contain channel information determined by the recording conditions, which is prone to over-fitting. For a more robust ASC system, We propose a robust feature learning (RFL) framework to train the CNN. The RFL framework down-weights CNN learning specifically on long-duration sounds. The proposed method is to train an auxiliary classifier with only long-duration sound information as input. The auxiliary classifier is trained with an auxiliary loss function that assigns less learning weight to poorly classified examples than the standard cross-entropy loss. The experimental results show that the proposed RFL framework can obtain a more robust acoustic scene classifier towards unseen devices and cities.

\end{abstract}


%
\IEEEpeerreviewmaketitle

\section{Introduction}
\label{sec:intro}

 

\IEEEPARstart{A}{coustic} scene classification (ASC) aims to identify the type of scene (environment) in which a given audio signal was recorded. Nowadays, wearable devices like smart-watches and smart bracelets are becoming popular. They could be equipped with microphones to capture sounds, e.g., for taking oral instructions from users. The microphones can also be used to collect environmental sounds that describe the acoustic scenes of recording. The acoustic scene signals can be analyzed to obtain context information. The context information enables the devices to make a more accurate response to user instructions. The hearing-impaired population may benefit from this technology with environment-aware acoustic information.

An ASC system generally comprises two major components: a feature extractor and a classifier. The feature extractor is used to extract time-frequency (TF) feature representation from raw audio waveforms. The classifier makes predictions based on the given TF feature representation. A TF feature representation can be viewed as an image with time and frequency being x-axis and y-axis, respectively. The value of an image corresponds to the signal intensity or power at a certain frequency and time. TF features can be obtained by the short-time Fourier transform (STFT), constant-Q transform \cite{CQT}, wavelet transform \cite{Ren2017}, log-mel filter-bank \cite{Wu2019,Hyeji2019,Huang2019}. In the Detection and Classification of
Acoustic Scenes and Events (DCASE) 2020 challenge, log-mel features are predominantly used to build ASC systems. For the classifier, early-stage ASC models include traditional machine learning algorithms such as the support vector machine (SVM) \cite{Roma2013}, Gaussian mixture model \cite{Yun2016} and i-vector \cite{Eghbal-Zadeh2016}. For deep learning algorithms, the multi-layer perceptron (MLP) \cite{Mun2016} and the recurrent neural network (RNN) \cite{Vu2016} are used. In recent years, the state-of-the-art performance is achieved by convolutional neural network (CNN). It has become the most popular classifier for ASC. In DCASE 2019, the top $5$ systems on task 1A are all based on CNN \cite{Chen2019, Koutini2019, Hyeji2019, Haocong2019, Huang2019}.

An acoustic scene signal generally comprises of various sounds overlapping in time and frequency. Some are representative sounds for acoustic scenes, while others may not be informative for identifying a specific scene. A typical scenario of sounds overlapping involves the long-duration sounds and the transient sounds. For example, in a park scene, there are bird-singing (transient) sounds and wind (long-duration) sound. Bird-singing sounds are prominent and common in natural parks. However, the wind sound's characteristics (e.g., the loudness, buzzing/rumbling) heavily depend on the weather and recording location. A CNN trained with the overlapping sounds may memorize the co-existing bird-singing sounds and the specific wind sound together as representative patterns for the park scene. In this case, given test audios containing only the bird-singing sounds, a trained CNN may fail to recognize the scene. This failure can be avoid when CNN is trained and tested on audios with individual sounds.

The long-duration sounds and the transient sounds differ in duration. To separate them, we proposed a TF feature decomposition method based on sound duration \cite{tfdecomp}. We here call it sound duration based decomposition (SDBD). Previous experiments have shown that CNN with the decomposed log-mel feature performs better than CNN with standard log-mel feature \cite{tfdecomp}. In this study, we further investigate the SDBD method on the suitable separation boundary for the long-duration sounds and short-duration sounds.

The robustness of ASC systems against unseen conditions is an important practical issue. By unseen condition, we mean that the recording condition is not covered in the training data but included in the test data. In real-world applications, acoustic scene signals may be recorded in unknown environments (e.g. in a city not involved in training data collection) or via devices not used in data collection. The mismatched characteristics between training data and test data expectedly lead to significant degradation of classification performance. 


Previous studies suggest that the long-duration sounds in acoustic scene signals, while carrying scene information, are prone to over-fitting in CNN model training. In \cite{Han2017}, a median filter was applied to remove background noise in TF features and improvement on the ASC accuracy on the TUT Acoustic Scenes 2017 dataset was reported \cite{Mesaros2016_EUSIPCO}. The processed features were said to be more robust against over-fitting, and make CNN models easier to detect sound events. This method is further interpreted and generalized in \cite{tfdecomp}. Experiment results on the TAU Urban Acoustic Scenes 2019 development dataset \cite{Mesaros2018_DCASE} show that jointly considering the long-duration sounds and transient sounds are better than considering transient sounds only. There appears to be a trade-off between making use of long-duration sound information and preventing CNN from over-fitting. 


The present study begins with an investigation on different definitions of the duration boundary that separate long-duration sounds and short-duration sounds. $4$ different long-short boundaries are experimented, i.e., $0.05$s, $0.25$s, $0.5$s, and $1$s. ASC accuracy is found to be increased with a longer long-short boundary. The difference between SDBD and the median filtering based HPSS method is discussed. Experimental results show that SDBD outperforms the HPSS method in multiple configurations. Based on decomposed TF feature, we propose a robust feature learning (RFL) framework that down-weights CNN learning specifically on long-duration sounds. We introduce an auxiliary classifier whose input is the time-averaged embedding feature of the component containing long-duration sounds. The auxiliary classifier is trained with an auxiliary loss function, which down-weights the learning of poorly classified training examples compared to the standard cross-entropy loss. Experiments are carried out under two scenarios, one is ASC with multiple (seen and unseen) devices, and another is with multiple (seen and unseen) cities. The results indicate that our proposed RFL framework can increase ASC systems' accuracy towards those unseen recording conditions. The main contributions of this paper are as follows: 
\begin{itemize}
    \item The duration boundary that separates long-duration sounds and short-duration sounds is studied.
    \item The importance of down-weighting CNN's embedding feature learning on long-duration sounds is addressed. 
    \item An RFL framework with novel loss functions is proposed to improve ASC system performance towards unseen recording conditions.
\end{itemize}

This paper is organized as follows. Section \ref{sec:datasets} introduces the datasets used for our experiments. Section \ref{sec:ASC_system} details the baseline ASC system. Section \ref{sec:tf_decomp} introduces the TF feature decomposition methods and the multi-input CNN architecture. Section \ref{sec:exp_decomposed_feats} studies the long-short boundary by experiments. Section \ref{sec:robust_feature_process} describes the feature processing methods to improve ASC robustness. Section \ref{sec:feature_learning_rfl} details the RFL framework for training more robust ASC systems. Section \ref{sec:experiments_robust_learn} shows the experiments on the feature processing methods and the proposed RFL framework. Section \ref{sec:conclusions} concludes our work.

\section{Dataset}
\label{sec:datasets}

Our experiments are conducted on $2$ datasets. The TAU Urban Acoustic Scenes 2020 Mobile development dataset \cite{dataset_asc_mobile_development_2020} is used to evaluate ASC system performance towards multiple (seen and unseen) recording devices. The TAU Urban Acoustic Scenes 2019 development dataset is used to evaluate the system performance towards multiple (seen and unseen) cities. For simplicity, we call them Multi-Device dataset and Multi-City dataset, respectively.

\subsection{Multi-Device Dataset}
\label{ssec:TAU2020}

The Multi-Device dataset is the TAU Urban Acoustic Scenes 2020 Mobile development dataset \cite{dataset_asc_mobile_development_2020}. It has been used in the DCASE 2020 challenge on ASC with multiple devices (subtask A). It consists of recordings from $10$ European cities using $9$ different devices. There are $3$ real devices (denoted as A, B, C) and $6$ simulated devices (denoted as S1-S6). Each audio recording is monaural, $10$-second long with a sampling rate of $44100$ Hz. There are $64$ hours of audio in the dataset, and most of the audios ($40$ hours) are from device A. 

Our experiments follow the officially provided train/test setup: the training set has $38.8$ hours of audios, and the test set has $8.25$ hours of audios. In this setup, audios from device S4-S6 only appear in the test set. With this dataset, ASC system performance towards both seen devices (A, B, C, S1-S3) and unseen devices (S4-S6) can be evaluated.

\subsection{Multi-City Dataset}
\label{ssec:TAU2019}

Audio recordings in the Multi-City dataset are from the TAU Urban Acoustic Scenes 2019 development dataset. They are used in the DCASE 2019 challenge on ASC (subtask A). There are $40$-hour audios from 10 different acoustic scene classes recorded with the same device (device A in Section \ref{ssec:TAU2020}). The audios were recorded in various locations at $10$ European cities. Each audio recording is binaural, $10$-second long with a sampling rate of $48000$ Hz. In our experiments, it is converted to monaural audio. Audios in this dataset may also appear in the Multi-Device dataset, in the form of re-sampled monaural audios.

In the officially provided train/test split, there is only one unseen city in the test set. To better validate ASC systems' performance with unseen cities, we manually design a train/test split with more unseen cities. The design is as follows. All audios from Vienna, Stockholm, and Prague are used as training data. For audios from Paris, Milan, Lyon, London, approximately half of them are selected as training data and another half as test data. All audios from Lisbon, Helsinki, and Barcelona are used as test data. As a result, the training set contains $7242$ audio recordings, and the test set contains $7158$ audio recordings. This dataset is used for the evaluation of ASC performance towards both seen cities and unseen cities.

\section{Segment-Based ASC System}
\label{sec:ASC_system}

An important consideration to build an ASC system is the choice of fundamental time unit of audio signal being processed by the acoustic scene classifier. Theoretically, an audio signal of short length (e.g., as short as $0.025$ second, a typical frame length for STFT) recorded in a scene is expected to contain information of that scene. However, for accurate classification, the acoustic scene signals should have longer length to accumulate more acoustic scene information.

There is a trade-off between ASC system accuracy and ASC system response speed. If we want a system with fast-response and low-latency, we may construct a classifier that predicts scene classes frame-by-frame (e.g., the typical frame length is $25$ ms for STFT based features). Suppose we want a system to have high accuracy. In that case, it should be designed so that it can integrate the audio information during an extended time period.

We build classifiers with input being audio segment ($1.28$-second long) rather than the complete $10$-second audio recording in the datasets. In real operation, the segment-based ASC system can predict acoustic scene immediately after receiving one audio segment input, or give a more accurate prediction after accumulating prediction results on multiple audio segments.

\subsection{General System Design}
\label{ssec:system_design}

\begin{figure}[!t]
\centering
\includegraphics[width=\linewidth]{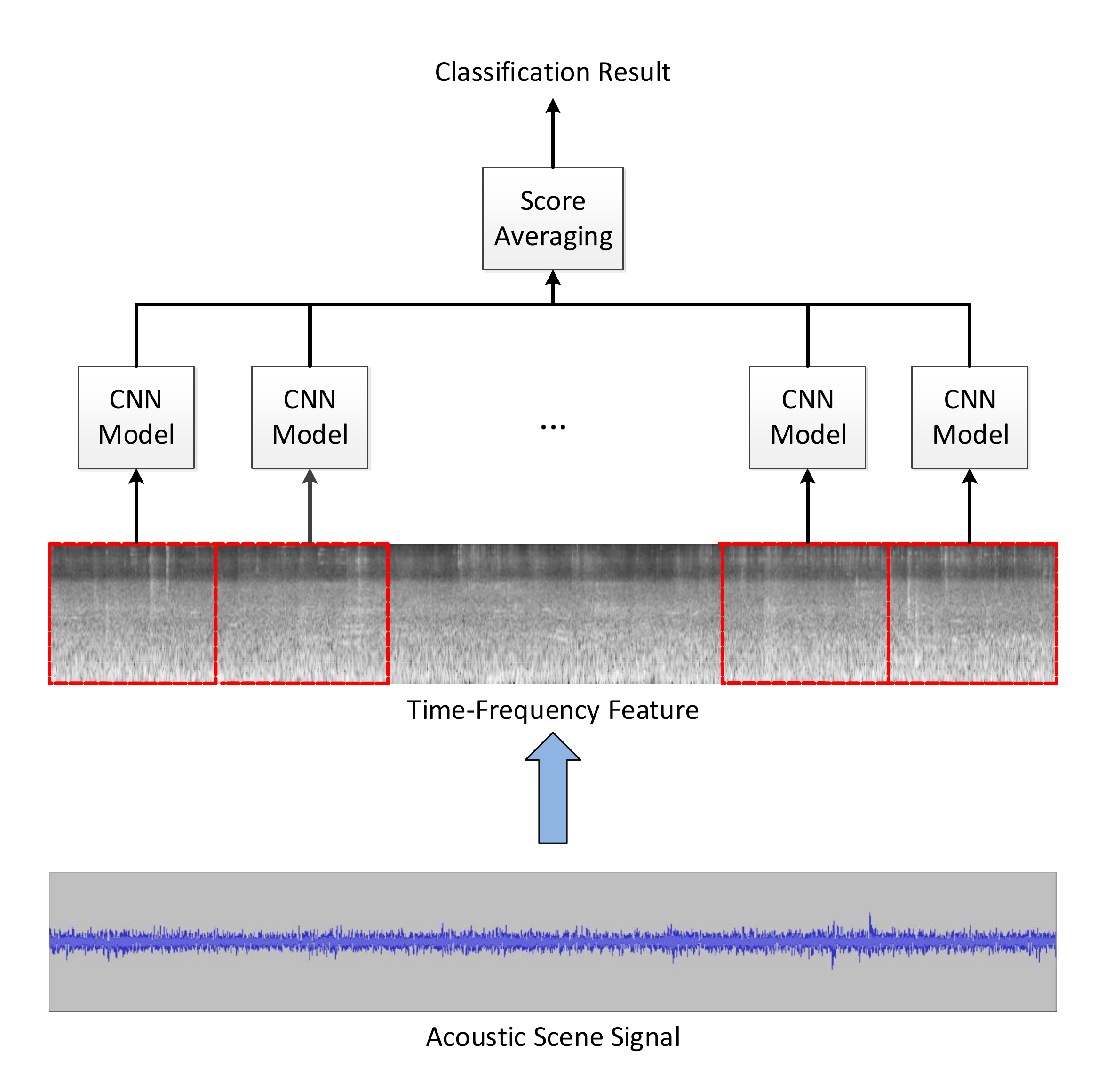}
\caption{General framework for segment-based ASC system. A TF feature is extracted from an acoustic scene signal. The TF feature is cut into fix-length segments for CNN to make predictions. The final classification score is the average of segment-level scores. }
\label{fig::seg_asc_framework}
\end{figure}

A general framework of segment-based ASC system is shown as in Figure \ref{fig::seg_asc_framework}. Given an acoustic scene signal, its TF representation is extracted. The TF feature is cut into non-overlapping feature segments. The CNN classifier gives a prediction of acoustic scenes for each TF feature segment. The prediction scores for feature segments are averaged to obtain the prediction scores of acoustic scenes for the signal. The scene class with the highest score is the predicted scene of the signal.

\subsection{Log-Mel Feature Extraction}
\label{ssec:feat_extract}



Among the TF feature representations for ASC, the log-mel feature is predominantly used. For each $10$-second audio signal in the datasets, STFT is applied on the audio waveform with $2048$ FFT points, window length of $25$ ms, and hop length of $10$ ms. The number of FFT points, window length, and hop length is empirically chosen. The window length is a typical length preferred for speech signal analysis \cite{prefer_20to40ms_window}. The hop length is decided such that approximately $100$ time frames represent $1$-second audio. The short time duration of the frame enables the CNN classifier to analyze short-duration sounds better. Log-mel filter-bank with $128$ filters is applied on the logarithm power of the STFT result. The obtained log-mel feature has the shape $(1000,128)$ where $1000$ is the number of time frames, and $128$ is the number of frequency bins. The number of frequency bins is chosen to be the same as the top-ranking ASC systems \cite{Hu2020, Chen2019} in DCASE challenges. 

Before a log-mel feature is fed into the CNN classifier, its values are normalized. This is referred to as the input data normalization \cite{input_normalization_importance}. The normalization is done for each frequency bin using the statistics of that frequency bin calculated from the whole training set. For normalized log-mel features in the training set, their values are in the range of $[-1,1]$.

\subsection{CNN Model}
\label{ssec:cnn}

The CNN architecture used in our experiments is shown as in Table \ref{table:model_structures}. $n$ denotes the number of input channels. For the baseline ASC system, the log-mel feature is used as input and thus $n=1$. From the top to the bottom of Table \ref{table:model_structures}, the CNN input, intermediate layers, and CNN output are described. ``3x3 Convolution-BN-ReLU ($48$ filters)'' means a stacking of a convolutional layer, a batch normalization (BN) layer, and a ReLU function. ``3x3'' is the kernel size, and ``$48$'' is the number of filters in the convolutional layer. The stride and padding are $1$ for convolutional layers. For max pooling layers, the padding is $0$. 

\begin{figure}[!t]
\centering
\includegraphics[width=0.9\linewidth]{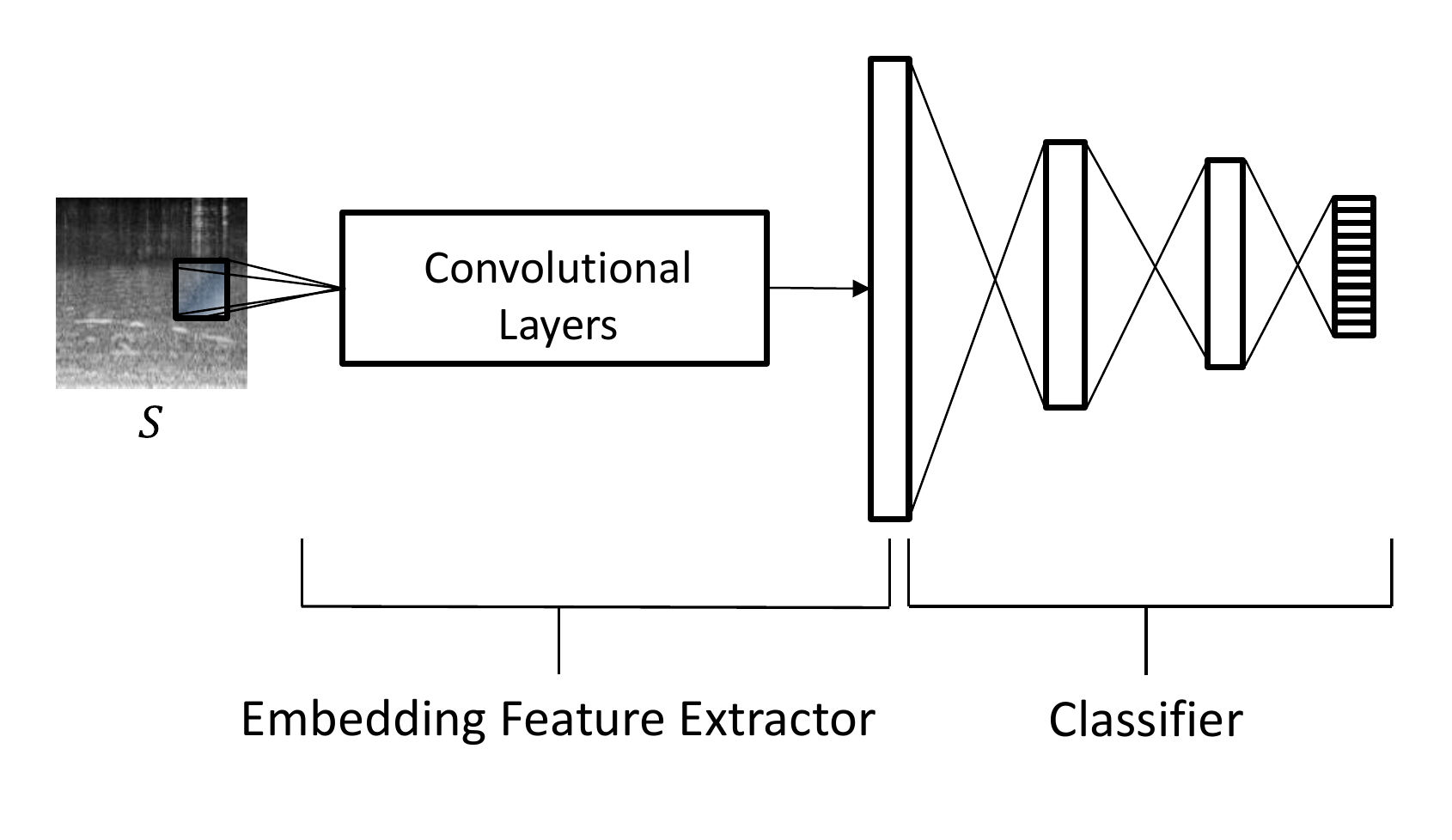}
\caption{An illustration of CNN viewed as a composition of a embedding feature extractor and a fully connected classifier. }
\label{fig::cnn_as_a_whole}
\end{figure}

The CNN can be viewed as the composition of $2$ parts: the convolutional part and the fully connected part, as shown in Figure \ref{fig::cnn_as_a_whole}. The convolutional part can be viewed as an embedding feature extractor, and the fully connected part can be viewed as a classifier. In the later sections, we propose to modify the training process of the embedding feature extractor to implement more robust ASC systems. 

To make a prediction on a $10$-second audio, the log-mel feature is extracted from the audio. It is cut into $1.28$-second non-overlapping feature segments. Thus, a $10$-second recording consists of $8$ segments. Each segment is fed into the CNN classifier to obtain the segment-level soft prediction scores. The overall prediction on the $10$-second audio recording is obtained by averaging the segment-level scores. 

\begin{table}
\centering
\caption{The CNN architecture used to construct our ASC system. $n$ is the number of input channels. }
\label{table:model_structures}
\begin{tabular}{cc} \bottomrule 
    1     & Input $n \times 128 \times 128$                       \\  \midrule
    2        & 3x3 Convolution-BN-ReLU ($48$ filters) \\
    3        & 2x2 Max Pooling                  \\
    4        & 3x3 Convolution-BN-ReLU ($96$ filters)       \\ 
    5        & 2x2 Max Pooling                 \\ 
    6        & 3x3 Convolution-BN-ReLU ($192$ filters)     \\
    7     & 2x2 Max Pooling                  \\ 
    8        & 3x3 Convolution-BN-ReLU ($192$ filters) \\
    9        & 3x3 Convolution-BN-ReLU ($192$ filters) \\
    10    & 2x2 Max Pooling                          \\ \midrule
    11    & Flattening                               \\
    12  & Fully Connected (dim-1024)-BN-ReLU    \\
    13  & Fully Connected (dim-256)-BN-ReLU     \\
    14    & 10-way Sigmoid                        \\\bottomrule
\end{tabular}
\end{table}

\section{TF Feature Decomposition}
\label{sec:tf_decomp}

Acoustic scene signal generally comprises of different types of sounds overlapping in time and frequency. As explained in Section \ref{sec:intro}, we expect that a separation of long-duration sounds and short-duration sounds could benefit ASC. This section begins with a review on median filtering and SDBD. The multi-input CNN is introduced to emphasize the independent embedding feature learning from long-duration sounds and short-duration sounds respectively. Lastly, we compare SDBD with the median-filtering based HPSS method. The HPSS method gives a soft separation of long-duration sounds and short-duration sounds while SDBD gives a hard separation of sounds.


\subsection{Median Filtering}
\label{ssec:median_filter}
Given an $1$-D discrete sequence $\mathbf{x}=[x_1,x_2,...,x_n]$ of length $n$, median filtering is applied with a moving window along the sequence. Let the window size of the median filter be $2k+1$ (k is a positive integer), the median-filtered value at time $t$, where $t \leq n$, is given by  
\begin{equation}
x_{\text{mf}}[t] = median([x_{t-k}, x_{t-k+1}, x_{t-k+2},...,x_{t+k}]),
\end{equation}
where $median()$ returns the median value of the input vector.

In image processing, the $2$-D median filter is commonly used to suppress impulse noise. The impulse noise is defined as undesirable high positive pixel values concentrated locally in a small region. On the other hand, there exist scenarios where impulse ``noise'' is desirable. In this case, subtracting the median-filtered signal from the original signal would retain impulse events that are narrower than half of the filtering window \cite{medfil_bs}. 

\subsection{Sound Duration Based Decomposition}
\label{ssec:sdbd}

The median filter was first proposed to process TF features for ASC in \cite{Han2017}, with the motivation of removing the steady background sound. Experiments were carried out in the TUT Acoustic Scenes 2017 dataset \cite{Mesaros2016_EUSIPCO}, and the ASC system using background removed TF feature showed better performance than the one using the original TF feature.  

However, removing the steady background sounds in the TF feature does not always result in higher ASC accuracy. In \cite{tfdecomp}, experiments were carried out in a larger dataset, showing that jointly considering the long-lasting sounds and transient sounds are better than considering transient sounds only. 


Instead of using median filtering for background sound removal in TF features \cite{Han2017}, our previous work \cite{tfdecomp} interpreted and generalized the usage of median filtering for ASC. Here we name our method as sound duration based decomposition (SDBD). The procedures of SDBD are as follows. Given a TF feature, median filtering is applied along the time axis on each frequency bin. After filtering, ``short'' impulse events (whose duration is less than half of the filter size) would be removed from the TF feature. Subtracting the filtered TF feature from the original one results in a feature of the same size that supposedly contains only those ``short'' events. As a result, the original TF feature is decomposed into two components, i.e., the median-filtered feature and the difference feature, which contain sounds of different duration ranges. 

For a separation of long-duration sounds and short-duration sounds, we set the number of components after SDBD to $2$. In this case, one median filter is used, and SDBD has a single parameter: the window size of that median filter. The procedures of SDBD to decompose a log-mel feature $S$ into $2$ components are stated as in Algorithm \ref{alg:medfil_strat_method}. Notice that by repeating the algorithm on $\mathbf{S_{long}}$ with an extra median filter, more components can be obtained. 

\renewcommand{\algorithmicensure}{\textbf{Procedure:}}
\begin{algorithm}[H] 
\caption{Applying SDBD on log-mel feature for separation of long-duration sounds and short-duration sounds. } 
\label{alg:medfil_strat_method} 
\begin{algorithmic}[1] 
\REQUIRE ~~\\ 
The original log-mel feature, $\mathbf{S}$;\\
Median filtering (along time axis) function $M_t(\cdot)$;
\ENSURE ~~\\ 
\STATE $\mathbf{S_{long}} = M_t(\mathbf{S})$; 
\label{ code:fram:get_notHigh }
\STATE $\mathbf{S_{short}} = \mathbf{S} - \mathbf{S_{long}}$; 
\label{code:fram:get_S_short}
\RETURN $( \mathbf{S_{long}}, \mathbf{S_{short}})$;
\end{algorithmic}
\end{algorithm}

SDBD has $2$ distinct properties. The first property is that summing up component features gives the original feature, i.e., $\mathbf{S_{long} + S_{short} = S}$. The second property is about its interpretability: given the frame length of the TF feature and the window size of the median filter, the exact range of ``long duration'' and ``short duration'' can be interpreted. For example, each frame represents $0.01$-second audio, and the median filter size is $201$. After decomposition, $S_{long}$ contains sounds whose duration is longer than half of the filter size, i.e., $100$ frames or $1$ second. $S_{short}$ contains sounds whose duration is shorter than $1$ second.   


\subsection{Multi-Input CNN}
\label{ssec:multi_input_cnn}

The characteristics between decomposed TF feature components are significantly different. After SDBD, in $S_{short}$ the sound patterns are present as vertical lines and noise of high spatial frequency. In contrast, sound patterns in $S_{long}$  are present as smoothly changing background. If the components were treated simply as a multi-channel image (similar to the $3$-channel RGB image in computer vision), the CNN would consider them independently only in the first convolutional layer. It would make little difference to CNN using the original TF feature as input, making the decomposition trivial. Thus, to emphasize the difference between components, independent embedding feature extractors are used for each type of component feature. The CNN using this design is called multi-input CNN. This idea mimics ensemble learning. The previous study has shown that it leads to a rise in classification accuracy \cite{tfdecomp}. 

\begin{figure}[!t]
\centering
\includegraphics[width=\linewidth]{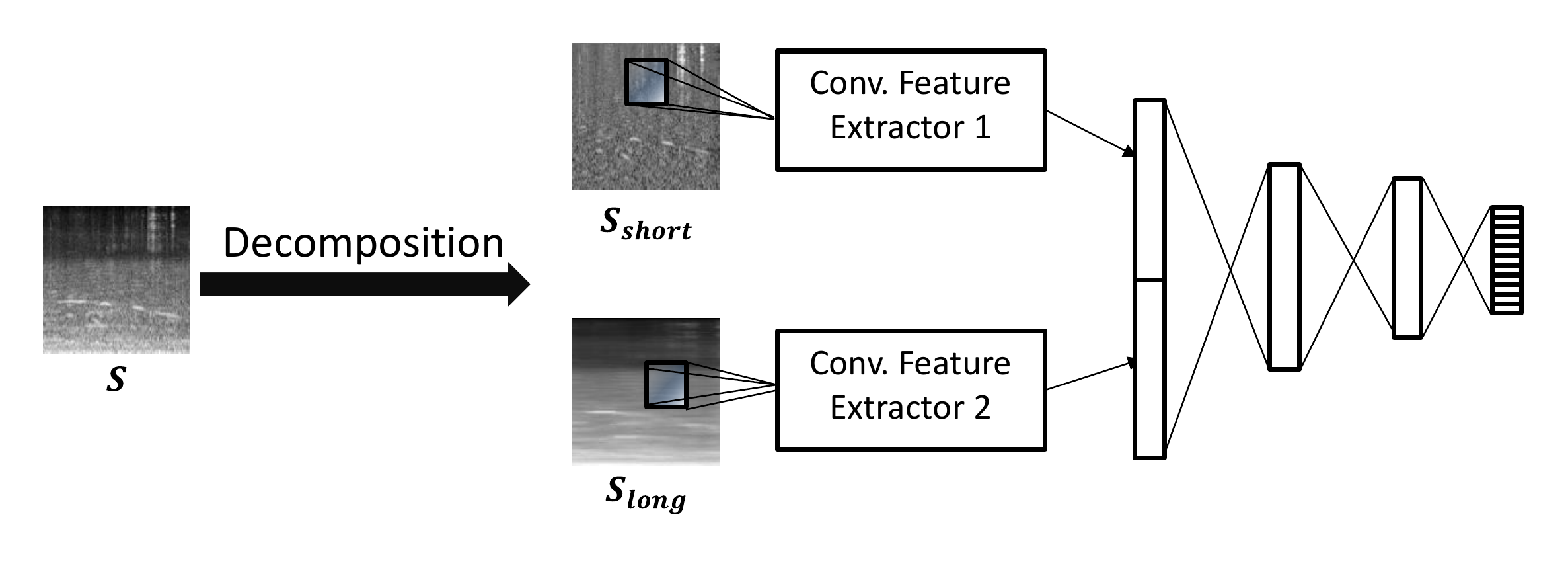}
\caption{An illustration of the multi-input CNN model with decomposed log-mel features for ASC. It consists of convolutional feature extractors for each component feature, and a fully connected classifier. }
\label{fig::mi_cnn}
\end{figure}

An illustration of the multi-input CNN is shown as in Figure \ref{fig::mi_cnn}. The input is the decomposed log-mel features using SDBD. Independent convolutional feature extractors are used for each type of component feature ($\mathbf{S_{long}}$ and $\mathbf{S_{short}}$). The extracted embedding features are concatenated as the input of the succeeding fully-connected classifier. In our experiments, multi-input CNN is implemented by splitting the convolution kernels in the benchmark CNN into $2$ groups, each receiving one type of component as input. It leads to fewer model parameters and faster computation speed.

\subsection{Median Filtering Based HPSS Method}
\label{ssec:hpss}

Harmonic-percussive source separation (HPSS) methods are first proposed for music signal processing. The goal of HPSS is to decompose an input audio signal into two signals: one consisting of all harmonic sounds and another consisting of all percussive sounds. It can also be regarded as to separate long-duration sounds and short-duration sounds, since the harmonic sounds are generally longer than the percussive sounds. 

There are many algorithms developed for HPSS. For example, in \cite{hpss_nmf_supervised}, non-negative matrix factorization (NMF) and SVM are used for HPSS. The method requires training data of harmonic signals and drum signals. In \cite{hpss_unsupervised}, a tensor factorization based algorithm is proposed without the need for prior knowledge of scores or included instruments.

In this study, we mainly discuss the median filtering based HPSS method \cite{hpss_median}. The method is based on the idea that percussive sounds can be regarded as vertical lines, and harmonic sounds as horizontal lines in a spectrogram. If not explicitly stated, ``the HPSS method'' refers to the median filtering based HPSS method in this study.

The procedures of the HPSS method are given as in Algorithm \ref{alg:hpss}. Given the power spectrogram $\mathbf{A}$ obtained by STFT, two median filters are applied along the time axis and frequency axis, respectively, to obtain two filtered spectrograms, $\mathbf{B_t}$ and $\mathbf{B_f}$ respectively. The filtered spectrograms are used as masks on $\mathbf{A}$ to generate the harmonic-enhanced spectrogram $\mathbf{H}$ and percussion-enhanced spectrogram $\mathbf{P}$. 

\renewcommand{\algorithmicensure}{\textbf{Procedure:}}
\begin{algorithm}[H] 
\caption{Applying the HPSS method on power spectrogram for separation of harmonic and percussive sounds. } 
\label{alg:hpss} 
\begin{algorithmic}[1] 
\REQUIRE ~~\\ 
The power spectrogram, $\mathbf{A}$;\\
Median filtering (along time axis) function $M_t(\cdot)$;\\
Median filtering (along frequency axis) function $M_f(\cdot)$;
\ENSURE ~~\\ 
\STATE $\mathbf{B_{t}} = M_t(\mathbf{A})$; 
\label{ code:fram:get_Bt }
\STATE $\mathbf{B_{f}} = M_f(\mathbf{A})$; 
\label{ code:fram:get_Bf }
\STATE $\mathbf{H} = \mathbf{B_{t}}/(\mathbf{B_{t} + B_{f}}) \cdot \mathbf{A}$; 
\label{ code:fram:get_H }
\STATE $\mathbf{P} = \mathbf{B_{f}}/(\mathbf{B_{t} + B_{f}}) \cdot \mathbf{A}$;  
\label{ code:fram:get_P }
\RETURN $(\mathbf{H}, \mathbf{P})$;
\end{algorithmic}
\end{algorithm}

Notice that ASC systems in this study take the log-mel feature as input. For a direct comparison of the HPSS method and SDBD, log-mel features are extracted from the decomposed spectrograms $\mathbf{H}$ and $\mathbf{P}$ using the configurations in Section \ref{ssec:feat_extract}.

\subsection{Comparison Between SDBD and the HPSS Method}
\label{ssec:compare_sdbd_hpss}

The HPSS method is similar to SDBD in terms of algorithmic procedures. SDBD applies median filtering along the time axis of TF features, while the HPSS method applies median filtering along both the time axis and frequency axis. 

The decomposition results of the HPSS method and SDBD are different. We call the HPSS method a soft separation method because of $2$ points. First, after HPSS each feature component still contains all the sounds in the original TF feature, with some of the sounds being suppressed rather than removed. Second, the duration boundary that separates long-duration sounds and short-duration sounds is not explicitly defined in the HPSS method. The SDBD is a hard separation method. In $S_{long}$, the short-duration sounds are removed. The duration boundary that separates long-duration and short-duration sounds is explicitly defined by a median filter size parameter. 


\section{Experiments on Decomposed TF Features}
\label{sec:exp_decomposed_feats}

In general, the parameters used in the HPSS method and SDBD are determined empirically. In this section, experiments are carried out to evaluate the ASC performance with TF feature decomposition using different parameter setups. The HPSS method and the SDBD are directly compared in terms of ASC accuracy. 


\subsection{Experimental Setup}
\label{ssec:exp_setup_tfdecomp}

For the HPSS method, a typical length for the harmonic filter $M_t$ and the percussive filter $M_f$ is $31$ \cite{hpss_autoencoder}. For SDBD, the boundary for short-duration and long-duration sounds varies from $0.25$ second to $1$ second. These setups were suggested in previous studies \cite{Han2017, tfdecomp}.

Experiments are carried out using $4$ different window sizes for median filters. The window sizes are $11$, $31$, $101$, and $201$. These values cover the typical values mentioned above. For the HPSS method, the window sizes of the harmonic filter and percussive filter are equal. In the SDBD case, the window sizes correspond to a long-short boundary of $0.05$ second, $0.15$ second, $0.50$ second, and $1$ second, respectively. By comparing the ASC performance under different setups, the preferred duration boundary that separates long-duration sounds and short-duration sounds can be found. 

\subsection{Model Training}
\label{ssec:model_train_tfdecomp}

The CNN models are implemented using PyTorch \cite{pytorch}. The hyper-parameters used for model training are the same for all experiments.  The initial learning rate is set to $0.0001$. The mini-batch size is $100$. The learning rate is multiplied by $0.5$ for every $4$ epochs. The number of training epochs is $40$. Model training is done by minimizing the binary CE loss with the Adam optimizer \cite{adam_optimizer} ($\beta _1 = 0.9$ and $\beta _2 = 0.999$). Weight decay with coefficient $0.0001$ is used for a regularization purpose. 

For data augmentation, the mixup \cite{mixup_imagenet} approach is used for each training batch. Given a batch of training samples $\{S_1,S_2,...,S_{100}\}$, each sample $S_i$ with label $y_i$ (which is an one-hot vector) is mixed with another randomly chosen sample $S_j$ with label $y_j$ in a random proportion $\lambda \in (0,1)$. The mixed training sample $S_{i}^{mix}$ and mixed label $y_{i}^{mix}$ is used for training:


\begin{equation}
\label{eq::mixed_sample}
\left.
\begin{aligned}
S_{i}^{mix} & = \lambda S_i + (1 - \lambda)S_j, \\
y_{i}^{mix} & = \lambda y_i + (1 - \lambda)y_j.  \\
\end{aligned}
\right.
\end{equation}

\subsection{Experimental Results}
\label{ssec:exp_results_tfdecomp}

Experiments are carried out on the Multi-Device dataset and the Multi-City dataset. The results are shown as in Table \ref{tab:acc_tfdecomps}.  The first column ``id'' indicates the unique index for each ASC system configuration. The second column ``ASC System'' describes the TF feature decomposition method used in the ASC system.  For example, ``Baseline'' is the baseline ASC system using the CNN model and the log-mel feature mentioned in Section \ref{sec:ASC_system}. ``SDBD (filter size $11$)'' means the multi-input CNN is used (following the implementation in Section \ref{ssec:multi_input_cnn}) with the decomposed log-mel features using SDBD. The window size of the median filter is $11$. ``HPSS (filter size $31$)'' means the multi-input CNN is used with the decomposed log-mel features using the HPSS method. The length of the median filters is $31$. 

The third column ``Long-Short Boundary'' refers to the duration boundary that separates long-duration sounds and short-duration sounds using the SDBD method. The boundary is equal to half of the median filter size multiplied with the STFT hop length (i.e., $0.01$s).

From column $4$ to column $6$, the accuracy of ASC systems trained and tested on the Multi-Device dataset is described. The accuracy is calculated based on the $10$-second audio recordings. Column $4$ shows the accuracy on audios from seen recording devices; column $5$ gives the accuracy on audios from unseen devices; column $6$ shows the overall accuracy on the Multi-Device dataset. From column $7$ to column $9$, the accuracy of ASC systems trained and tested on the Multi-City dataset is described. The highest accuracies are in bold.

\renewcommand{\arraystretch}{} 
\begin{table*}[ht]  
  \centering  
  \begin{threeparttable}  
  \caption{Accuracy of multi-input CNNs with different TF feature decomposition methods. }  
  \label{tab:acc_tfdecomps}  
    \begin{tabular}{llccccccc}  
    \toprule  
     & & & \multicolumn{3}{c}{Multi-Device Dataset}&\multicolumn{3}{c}{Multi-City Dataset}\cr  
    \cmidrule(lr){4-6} \cmidrule(lr){7-9}
    id &ASC System & Long-Short Boundary &Seen Device &Unseen Device &Overall &Seen City &Unseen City &Overall \cr  
    \midrule  
    1 & Baseline                  & - &$63.2\%$ &$45.5\%$ &$57.3\%$ &$71.6\%$ &$55.3\%$ &$61.8\%$\cr  
    2 & SDBD (filter size $11$) &$0.05$s &$62.1\%$ &$45.5\%$ &$56.6\%$ &$69.9\%$ &$56.2\%$ &$61.7\%$\cr  
    3 & SDBD (filter size $31$) &$0.15$s &$\mathbf{65.8\%}$ &$47.3\%$ &$59.7\%$ &$\mathbf{72.7\%}$ &$\mathbf{57.1\%}$ &$\mathbf{63.3\%}$\cr
    4 & SDBD (filter size $101$)  &$0.50$s &$64.6\%$ &$\mathbf{49.9\%}$ &$59.7\%$ &$72.3\%$ &$56.9\%$ &$63.0\%$\cr  
    5 & SDBD (filter size $201$)  &$1.00$s &$65.6\%$ &$49.7\%$ &$\mathbf{60.3\%}$ &$72.6\%$ &$56.5\%$ &$62.9\%$\cr  
    6 & HPSS (filter size $11$)   & - &$62.1\%$ &$44.6\%$ &$56.3\%$ &$71.6\%$ &$56.2\%$ &$62.3\%$\cr  
    7 & HPSS (filter size $31$)   & - &$62.6\%$ &$46.4\%$ &$57.2\%$ &$70.8\%$ &$55.9\%$ &$61.8\%$\cr  
    8 & HPSS (filter size $101$)  & - &$63.0\%$ &$44.8\%$ &$56.9\%$ &$72.3\%$ &$55.2\%$ &$62.0\%$\cr  
    9 & HPSS (filter size $201$)  & - &$62.0\%$ &$44.0\%$ &$56.0\%$ &$71.7\%$ &$55.0\%$ &$61.7\%$\cr  
    \bottomrule  
    \end{tabular}  
    \end{threeparttable}  
\end{table*}

\subsection{Discussion}
\label{ssec:discussion_tfdecomp}

In the Multi-Device dataset, the decomposed log-mel features using SDBD are better than the undecomposed log-mel features, if the duration boundary is larger than or equal to $0.15$s. The increase of long-short boundary from $0.05$s to $1$s leads to increasing overall accuracy. We think this trend is because of the better separation of transient sounds and long-duration sounds. The long-duration sounds carry channel information which varies with recording devices. 
Thus, with a longer boundary ($>1$s), a further performance gain could be expected, at the cost of larger computation cost. On the other hand, if the long-short boundary is too short (e.g., $0.05$ second), it will be shorter than the duration of a transient sound. In this case, SDBD will break the acoustic structure in a transient sound and thus lead to performance degradation. 

The system 5 (SDBD with median filter size $201$) achieves the best performance in the Multi-Device dataset. The system's overall accuracy is increased by $3.0\%$, and the accuracy towards audio from unseen devices is increased by $4.2\%$ compared to the baseline system. The system is also better than the baseline in the Multi-City dataset. 

In the Multi-City dataset, the SDBD method performs better than the baseline when the long-short boundary is larger than or equal to $0.15$s. Systems with SDBD's long-short boundary being $0.15$s, $0.5$s, and $1$s perform similarly well. It is different from the observed trend in the Multi-Device dataset. The reason could be that audios in this dataset are from a single recording device, and thus the effect of channel information is smaller than the Multi-Device dataset. 

All systems using the HPSS method (system 6-9) have no significant performance gain compared than the baseline system. In most of the cases, the HPSS method leads to a worse accuracy. The reason could be that the HPSS method is a soft separation method. The device-sensitive (and location-sensitive) channel information is not separated into $S_H$, but is distributed in both feature components.

To summarize, the following conclusions are made:
\begin{itemize}
\item In general, ASC systems using the SDBD outperform the baseline system when the long-short boundary is larger than or equal to $0.15$ second.
\item For ASC with unseen recording devices, a large duration boundary of SDBD is preferred (e.g., $1$ second).
\item The HPSS method does not help increase the ASC system accuracy significantly. 
\end{itemize}


\section{Feature Processing for More Robust ASC}
\label{sec:robust_feature_process}

In this section, we discuss several feature processing methods that can obtain more robust feature representations for ASC. The methods are based on the idea that the long-duration sounds contain channel information that is highly specific to recording devices and locations, and is not generalizable. These methods remove the channel information at the cost of losing scene information. 


\subsection{Log-Spectral Mean Normalization}
\label{ssec:LSMN}

The log-spectral mean normalization (LSMN) aims to remove channel effects on log-magnitude spectral features. The channel effects are caused by the transmission systems. Let $T$ be the total number of frames, and $x_i[n]$ denote the $i$-th frame of the input signal. Assume the channel is a linear time-invariant (LTI) system and the impulse response is $h[n]$, the recorded signal $y_i[n]$ is given by convolving $h[n]$ on $x_i[n]$:
\begin{equation}
\label{eq:time_domain_channel_effect}
y_i[n] = h[n] \star x_i[n].
\end{equation}
Taking Fourier transform and applying logarithm to the magnitude spectrum, we have 
\begin{equation}
\label{eq:logPS_domain_channel_effect}
log(|Y_i[f]|) = log(|H[f]|) + log(|X_i[f]|).  
\end{equation}

Subtracting $log(|Y_i[f]|)$ by its mean over the $T$ frames gives the result of LSMN, and the channel effect term $H[f]$ is removed: 
\begin{equation}
\label{eq::lsmn_result}
R_i[f] = log(|X_i[f]|) - \frac{1}{T}\sum^{T}_{i=1}{log(|X_i[f]|)}.
\end{equation}

LSMN adopts the same idea behind cepstral mean normalization (CMN). The name LSMN is used because we are handling the features in log-spectral domain. CMN is applied to cepstral features, e.g., mel frequency cepstrum coefficients (MFCC). CMN is a well-established technique for speaker verification \cite{cepstral_analysis_sv} and speech recognition \cite{cepstral_asr}. It assumes a linear channel distortion in the time domain, which leads to a constant offset in the cepstral domain. 

\subsection{Discarding Long-Duration Sounds in TF Feature}
\label{ssec:discard_Slong}

Previous studies \cite{Han2017,enhanceSoundTexture} suggest that the slow-varying background sounds can be sensitive to the change of recording conditions, they are prone to over-fitting in CNN model training. In the previous section, we describe the feature decomposition methods (HPSS and SDBD) to separate long-duration sounds and short-duration sounds. After decomposition, the long-duration feature component can be discarded.

With the HPSS method, we can obtain a harmonic-enhanced log-mel feature $S_H$ and a percussion-enhanced log-mel feature $S_P$. Since the harmonic sounds are longer than the percussive sounds and are suppressed in $S_P$, discarding $S_H$ and only using $S_P$ as the input feature for ASC systems achieves the purpose of suppressing long-duration sounds.

With SDBD, decomposing log-mel feature results in $2$ feature components, i.e., $S_{short}$ and $S_{long}$. Discarding $S_{long}$ and only using $S_{short}$ as the input of ASC systems achieves the goal of discarding long-duration sounds.

It should be noted that long-duration sounds also contain acoustic scene information that is not dependent on specific recording conditions. Thus, completely discarding the long-duration sounds may not be the best choice. A better approach is presented in Section \ref{sec:feature_learning_rfl}.

\section{Feature Learning For More Robust ASC}
\label{sec:feature_learning_rfl}

Though completely discarding long-duration sounds in an acoustic scene would be helpful to improve the robustness of ASC systems to a certain extent, it inevitably leads to the loss of acoustic scene information. Ideally, we hope to make the system more robust without losing useful scene information. An alternative method is to suppress the acoustic information being learned from long-duration sounds, or say to down-weight the CNN learning on such sounds. In this way, the CNN can learn the scene-related information from long-duration sounds, which does not depend on recording conditions, and thus the ASC system performance towards unseen recording conditions is improved. We propose a robust feature learning (RFL) framework to train multi-input CNNs with decomposed TF features.

\begin{figure}[!t]
\centering
\includegraphics[width=\linewidth]{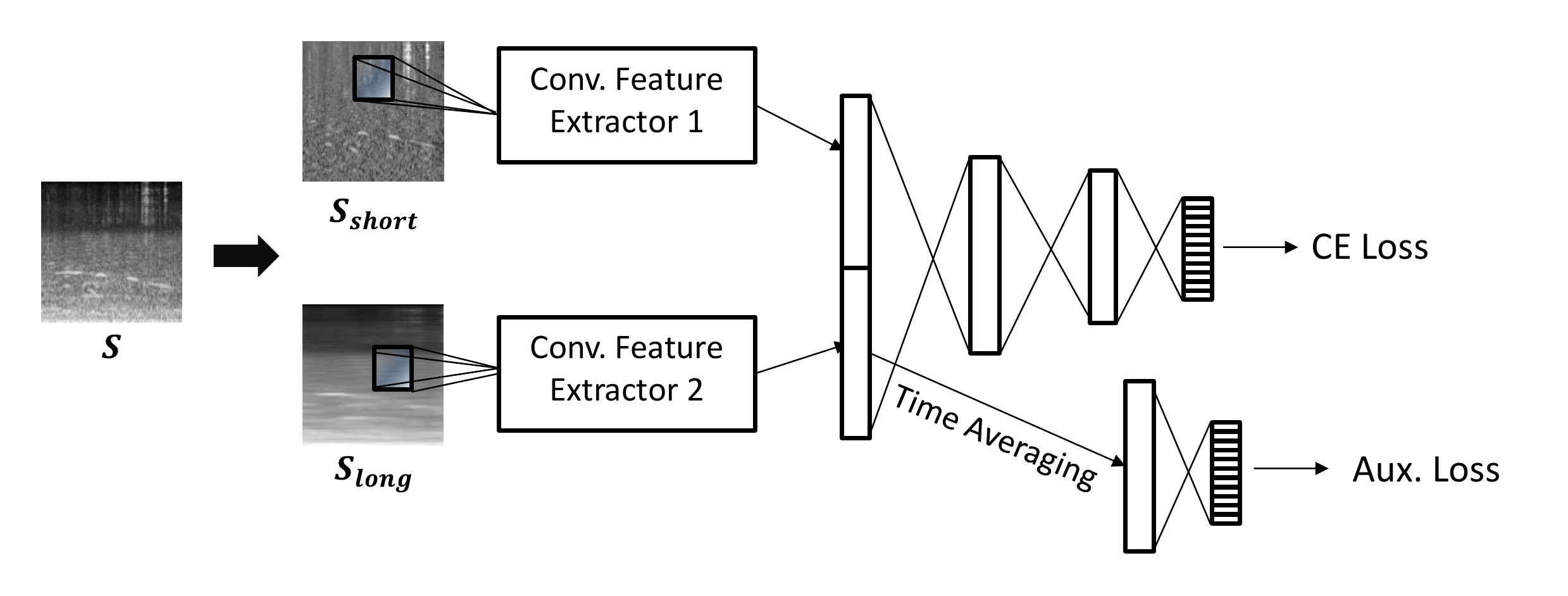}
\caption{An illustration of the RFL framework for training multi-input CNN with decomposed TF features. An auxiliary classifier is introduced in the training stage. The input of the auxiliary classifier is the time-averaged embedding feature extracted from long-duration sounds. The classifier is trained with an auxiliary loss that assigns less learning weight to poorly classified examples than the CE loss. }
\label{fig::cnn_LS_auxiliary}
\end{figure}

An illustration of the RFL framework that aims at down-weighting CNN learning on long-duration sounds is given as in Figure \ref{fig::cnn_LS_auxiliary}. It is based on the multi-input CNN model described in Section \ref{ssec:multi_input_cnn}. After the SDBD, the original log-mel feature $S$ is decomposed into $S_{short}$ and $S_{long}$. We introduce an auxiliary classifier, whose input is the time-averaged embedding feature extracted from $S_{long}$. The classifier is trained with an auxiliary loss function that leads to a smaller learning weight to poorly classified examples than the standard cross-entropy (CE) loss. During model training, the overall loss for back-propagation is the sum of the CE loss for the primary classifier and the loss for the auxiliary classifier. The auxiliary classifier is discarded in the testing stage. No extra model parameter is introduced in the trained CNN model.

\subsection{Cross-Entropy Loss for CNN Training}
\label{ssec:ce_loss}

\begin{figure}[!t]
\centering
\subfloat[The CE loss with respect to output probability. ]{\includegraphics[width=\columnwidth]{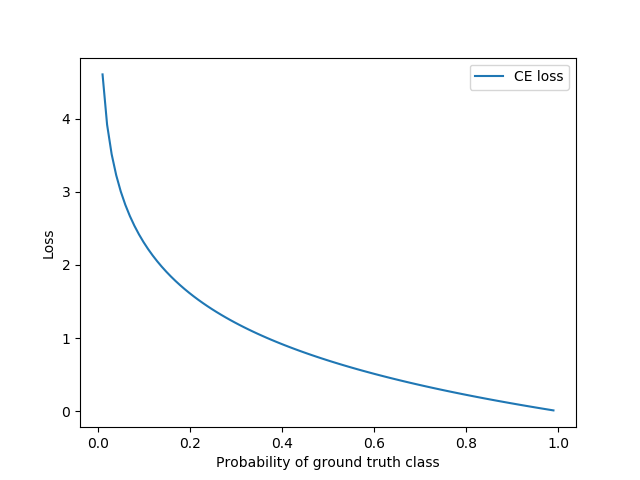}%
\label{fig::ce_loss_p}}
\hfil
\subfloat[Gradient of the CE loss with respect to logit. ]{\includegraphics[width=\columnwidth]{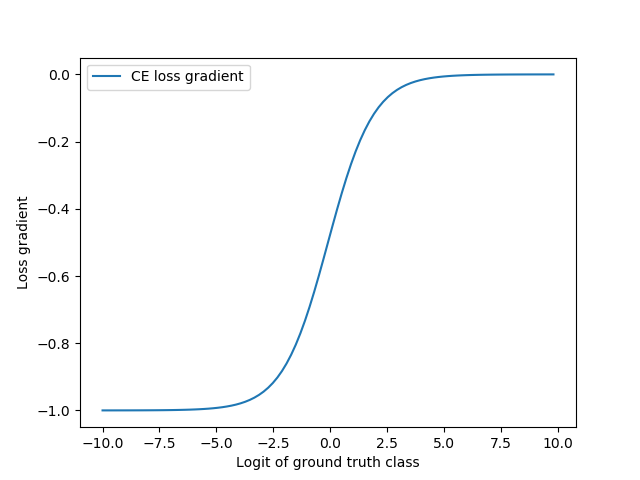}%
\label{fig::ce_gloss_logit}}
\caption{An illustration of the CE loss. It can be seen from (b) that the value of gradient w.r.t to logit lies in the interval of $(-1,0)$.}
\label{fig::ce_loss}
\end{figure}

The expression of the CE loss function for ground-truth class is $L_{CE}(p) = - log(p)$, where $p$ is the output probability for the ground-truth class. Figure \ref{fig::ce_loss} illustrate the (a) change of CE loss with respect to output probability of ground-truth class and (b) its gradient with respect to logit (i.e., the value before applying the sigmoid function). According to Figure \ref{fig::ce_loss_p}, the curve is steeper in low probability region, indicating a larger gradient magnitude for back-propagation. The gradients can be seen more clearly in Figure \ref{fig::ce_gloss_logit}, whose value is always negative and lies in the range of $(-1,0)$. 

We emphasize that it is the gradient magnitude of the loss function that decides how the model parameters are updated. A large magnitude indicates an aggressive update (a large learning weight), while a small magnitude leads to a conservative/mild update (a small learning weight). With the CE loss, the poorly classified examples (i.e., examples with low probability for ground-truth class) will be updated more aggressively than the well-classified examples (with high probability for ground-truth class).

\subsection{Auxiliary Loss Functions}
\label{ssec:aux_losses}

The arrangement of gradient magnitudes (learning weights) towards training examples in the CE loss may not always be optimal. Long-duration sounds contain many acoustic scene patterns that are dependent on recording devices and recording locations, while short-duration sounds' characteristics are more robust against recording conditions. A smaller learning weight on long-duration sounds can limit the amount of learned acoustic patterns that are dependent to recording conditions. Thus, we propose two auxiliary loss functions that have smaller gradient magnitude for poorly classified examples than the CE loss, i.e., the defocus loss and the reversed cross-entropy (RCE) loss.

\subsubsection{Defocus Loss}
\label{sssec:defoucs_loss}

The defocus loss mimics the focal loss \cite{focalloss} in an opposite manner. The focal loss emphasizes more on poorly classified examples than the CE loss by adjusting the steepness of the CE loss curve. The defocus loss tends to de-emphasize poorly classified examples. Denote $p \in (0,1) $ as the output probability of the CNN classifier. The defocus loss for ground-truth class is given by 
\begin{equation}
\label{eq::defocus_loss}
L_{df}(p) = - e^{\alpha p} log(p),
\end{equation}
where $\alpha \in [0,1]$ is a parameter controlling the degree of focus on poorly classified examples. When $\alpha = 0$, the loss is identical to the CE loss. With $\alpha$ increasing, the learning weight on poorly classified examples will decrease.  

\subsubsection{Reversed Cross-Entropy (RCE) Loss}
\label{sssec:reversed_CELoss}


To enable wider range of adjustment in the ``degree of focus on poorly classified examples'' than the defocus loss, we propose the reversed cross-entropy (RCE) loss. The RCE loss is designed such that it has a gradient curve (with respect to logit) emulating a reversed gradient curve of the CE loss. The RCE loss puts less focus on poorly classified training examples. Denote $p \in (0,1) $ as the output probability of the CNN classifier. The reverse loss is given by 
\begin{equation}
\label{eq::loss_rce}
L_{R}(p) = log(1-p). 
\end{equation}
The RCE loss is a weighted sum of the CE loss and the reverse loss. Let $\alpha \in [0,1]$ be the weight of the reverse loss. The RCE loss is given by 
\begin{equation}
\label{eq::loss_mix_ce_rce}
L_{RCE}(p) = (1-\alpha )L_{CE}(p) + \alpha L_{R}(p).
\end{equation}
The degree of focus on poorly classified examples can be adjusted by tuning $\alpha$. A larger $\alpha$ indicates less focus on these examples. When $\alpha = 0.5$, the RCE loss w.r.t. logit (i.e., the value before applying the sigmoid function) $L_{RCE}(x)$ becomes a linear function. Though there is no lower bound for the value of the RCE loss, in practice it is observed that the RCE loss becomes stable as model training proceeds.


\subsection{Comparing CE Loss and Auxiliary Loss}
\label{ssec:compare_ce_aux}

\begin{figure}[!t]
\centering
\subfloat[The losses with respect to output probability. ]{\includegraphics[width=\columnwidth]{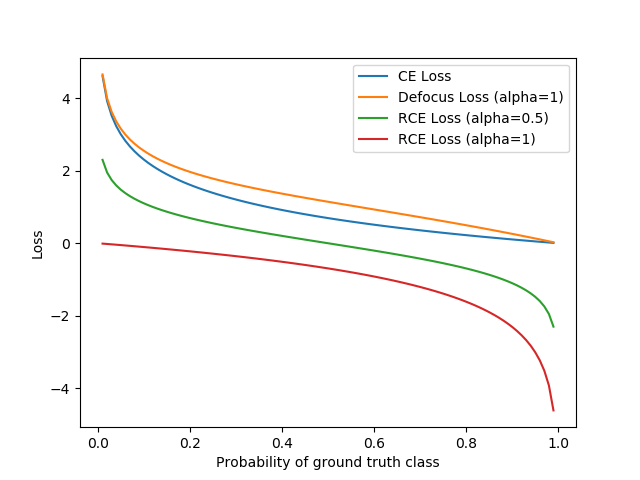}%
\label{fig::losses_p}}
\hfil
\subfloat[Gradients of the losses with respect to logit. ]{\includegraphics[width=\columnwidth]{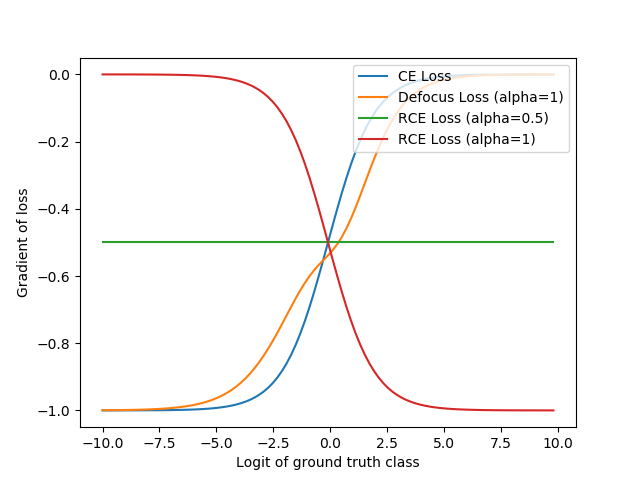}%
\label{fig::losses_gradient_logit}}
\caption{An illustration of the CE loss and the auxiliary losses (i.e., the defocus loss and the RCE loss). By proper selection of the auxiliary loss function, we may obtain the optimal degree of focus on poorly classified examples. }
\label{fig::losses}
\end{figure}

In Figure \ref{fig::losses}, the CE loss and the auxiliary losses (i.e., the defocus loss and the RCE loss) are shown for comparison. The degree of focus on poorly classified examples for different loss functions can be viewed clearly in Figure \ref{fig::losses_gradient_logit}. According to the gradient magnitude in the low probability region (negative-valued logit), it can be seen that the CE loss has the strongest focus on poorly classified examples. The defocus loss ($\alpha = 1$) takes second place. The RCE loss with $\alpha = 0.5$ takes third place. The RCE loss with $\alpha = 1$ has the least focus on poorly classified examples. With proper selection of auxiliary function, the degree of focus on poorly classified examples can be adjusted flexibly.

\section{Experiments on Robust ASC}
\label{sec:experiments_robust_learn}

In this section, experiments are carried out to compare the efficacy of the feature processing methods described in Section \ref{sec:robust_feature_process} and the RFL framework proposed in Section \ref{sec:feature_learning_rfl} that aims to increase ASC systems' robustness towards unseen recording conditions. 

\subsection{Experimental Setup}
\label{ssec:experimental_setup_robustLearn}

For the methods utilizing the decomposed TF features, the setups for the TF feature decomposition methods are based on the experimental conclusions in Section \ref{ssec:discussion_tfdecomp}, i.e., the median filer size is $31$, $101$, and $201$, respectively.

For experiments on the feature processing methods, the LSMN method (in Section \ref{ssec:LSMN}) and the discarding long-duration sounds (DLDS) approach (in Section \ref{ssec:discard_Slong}) are used. The DLDS approach is applied on log-mel features decomposed with the HPSS method and the SDBD method. 

For experiments on the RFL framework, model training is done by decreasing the binary overall loss. The overall loss is a sum of the CE loss on the main classifier and the auxiliary loss on the auxiliary classifier. Multiple auxiliary loss functions (described in Section \ref{ssec:compare_ce_aux}) are experimented. 

\subsection{Experimental Results}
\label{ssec:exp_results_robustLearn}

\renewcommand{\arraystretch}{} 
\begin{table*}[ht]  
  \centering  
  \begin{threeparttable}  
  \caption{Accuracy of CNNs with various feature prepocessing methods, and CNNs trained with the RFL framework. }  
  \label{tab:acc_robust_learn}  
    \begin{tabular}{llcccccc}  
    \toprule  
    & &\multicolumn{3}{c}{Multi-Device Dataset} &\multicolumn{3}{c}{Multi-City Dataset}\cr  
     \cmidrule(lr){3-5} \cmidrule(lr){6-8}
    id   & ASC System &Seen Device &Unseen Device &Overall &Seen City &Unseen City &Overall \cr  
    \midrule  
    1  & Baseline                                                   &$63.2\%$ &$45.5\%$ &$57.3\%$ &$71.6\%$ &$55.3\%$ &$61.8\%$\cr 
    10  & Log-spectral mean normalization                                &$65.0\%$ &$58.7\%$ &$62.9\%$ &$65.3\%$ &$56.7\%$ &$60.1\%$\cr 
    \midrule 
    11  & HPSS (filter size 31) + DLDS                               &$59.8\%$ &$39.0\%$ &$52.9\%$ &$70.6\%$ &$57.1\%$ &$62.4\%$\cr  
    12  & HPSS (filter size 101) + DLDS                              &$63.0\%$ &$42.7\%$ &$56.3\%$ &$71.2\%$ &$58.3\%$ &$63.4\%$\cr  
    13  & HPSS (filter size 201) + DLDS                              &$62.4\%$ &$46.5\%$ &$57.1\%$ &$71.3\%$ &$57.8\%$ &$63.2\%$\cr  
    14  & HPSS (filter size 31) + RFL (defocus loss $\alpha=1$)      &$62.1\%$ &$45.3\%$ &$56.5\%$ &$72.7\%$ &$56.9\%$ &$63.2\%$\cr  
    15  & HPSS (filter size 101) + RFL (defocus loss $\alpha=1$)     &$61.9\%$ &$44.9\%$ &$56.2\%$ &$72.5\%$ &$57.1\%$ &$63.2\%$\cr  
    16  & HPSS (filter size 201) + RFL (defocus loss $\alpha=1$)     &$62.5\%$ &$44.2\%$ &$56.4\%$ &$71.6\%$ &$55.0\%$ &$61.6\%$\cr  
    17  & HPSS (filter size 31) + RFL (RCE loss $\alpha=0.5$)        &$60.2\%$ &$40.3\%$ &$53.6\%$ &$69.9\%$ &$56.8\%$ &$62.0\%$\cr  
    18 & HPSS (filter size 101) + RFL (RCE loss $\alpha=0.5$)       &$61.0\%$ &$44.0\%$ &$55.3\%$ &$68.5\%$ &$56.1\%$ &$61.1\%$\cr  
    19 & HPSS (filter size 201) + RFL (RCE loss $\alpha=0.5$)       &$60.2\%$ &$44.9\%$ &$55.1\%$ &$69.3\%$ &$56.3\%$ &$61.5\%$\cr  
    20 & HPSS (filter size 31) + RFL (RCE loss $\alpha=1$)          &$58.3\%$ &$38.6\%$ &$51.8\%$ &$69.1\%$ &$56.7\%$ &$61.7\%$\cr  
    21 & HPSS (filter size 101) + RFL (RCE loss $\alpha=1$)         &$60.5\%$ &$44.2\%$ &$55.1\%$ &$70.3\%$ &$56.4\%$ &$61.9\%$\cr  
    22 & HPSS (filter size 201) + RFL (RCE loss $\alpha=1$)         &$61.3\%$ &$45.4\%$ &$56.0\%$ &$70.3\%$ &$57.0\%$ &$62.3\%$\cr  
    \midrule 
    23 & SDBD (filter size 31) + DLDS                               &$46.4\%$ &$46.5\%$ &$46.4\%$ &$63.8\%$ &$53.7\%$ &$57.7\%$\cr  
    24 & SDBD (filter size 101) + DLDS                              &$62.5\%$ &$59.7\%$ &$61.5\%$ &$65.1\%$ &$56.4\%$ &$59.8\%$\cr  
    25 & SDBD (filter size 201) + DLDS                              &$63.0\%$ &$61.9\%$ &$62.6\%$ &$65.1\%$ &$56.1\%$ &$59.6\%$\cr  
    26 & SDBD (filter size 31) + RFL (defocus loss $\alpha=1$)      &$64.5\%$ &$48.6\%$ &$59.2\%$ &$71.9\%$ &$57.7\%$ &$63.3\%$\cr  
    27 & SDBD (filter size 101) + RFL (defocus loss $\alpha=1$)     &$65.5\%$ &$49.7\%$ &$60.3\%$ &$\mathbf{73.6\%}$ &$56.1\%$ &$63.0\%$\cr  
    28 & SDBD (filter size 201) + RFL (defocus loss $\alpha=1$)     &$\mathbf{66.1\%}$ &$50.0\%$ &$60.7\%$ &$70.7\%$ &$56.1\%$ &$61.9\%$\cr  
    29 & SDBD (filter size 31) + RFL (RCE loss $\alpha=0.5$)        &$60.5\%$ &$55.9\%$ &$59.0\%$ &$61.2\%$ &$53.5\%$ &$56.6\%$\cr  
    30 & SDBD (filter size 101) + RFL (RCE loss $\alpha=0.5$)       &$63.7\%$ &$60.0\%$ &$62.5\%$ &$70.2\%$ &$58.1\%$ &$63.0\%$\cr  
    31 & SDBD (filter size 201) + RFL (RCE loss $\alpha=0.5$)       &$64.3\%$ &$61.8\%$ &$63.5\%$ &$70.6\%$ &$\mathbf{59.3\%}$ &$\mathbf{63.8\%}$\cr  
    32 & SDBD (filter size 31) + RFL (RCE loss $\alpha=1$)          &$54.2\%$ &$51.5\%$ &$53.3\%$ &$66.1\%$ &$58.5\%$ &$61.6\%$\cr  
    33 & SDBD (filter size 101) + RFL (RCE loss $\alpha=1$)         &$63.8\%$ &$61.7\%$ &$63.1\%$ &$71.4\%$ &$57.5\%$ &$63.0\%$\cr  
    34 & SDBD (filter size 201) + RFL (RCE loss $\alpha=1$)         &$64.4\%$ &$\mathbf{62.2\%}$ &$\mathbf{63.7\%}$ &$70.9\%$ &$58.7\%$ &$63.6\%$\cr 
    
    \bottomrule  
    \end{tabular}  
    \end{threeparttable}  
\end{table*}

Experiments are carried out in both the Multi-Device dataset and the Multi-City dataset. The experimental results are shown as in Table \ref{tab:acc_robust_learn}. The first column ``id'' indicates the unique index for each ASC system configuration.

The second column ``ASC System'' describes the feature processing methods or the RFL framework used in the ASC system. For example, ``Baseline'' is the baseline ASC system using the CNN model and the log-mel feature mentioned in Section \ref{sec:ASC_system}.  ``Log-spectral mean normalization'' means the LSMN method is used to process the log-mel features. ``HPSS (filter size 31) + DLDS'' means the CNN uses the harmonic-enhanced log-mel features as input, which are obtained by the HPSS method (filter size being $31$). The percussive-enhanced features are discarded based on the DLDS approach. ``SDBD (filter size $201$) + DLDS'' means the CNN uses $S_{short}$ as input, which are obtained by the SDBD method (filter size being $201$). $S_{long}$ is discarded based on the DLDS approach. Similarly, ``SDBD (filter size 201) + RFL (RCE loss $\alpha=1$)'' means the multi-input CNN is trained with the RFL framework, with auxiliary loss function being the RCE loss ($\alpha = 1$). The input of the CNN is log-mel features decomposed with the SDBD method (with median filter size being $201$).

From column $3$ to column $5$, the accuracy of ASC systems trained and tested on the Multi-Device dataset is described. Column $3$ shows the accuracy on audios from seen recording devices; column $4$ gives the accuracy on audios from unseen devices; column $5$ shows the overall accuracy on the Multi-Device dataset. From column $6$ to column $8$, the accuracy of ASC systems trained and tested on the Multi-City dataset is given. The accuracy is calculated based on the $10$-second audio recordings. The highest accuracies are in bold.

\subsection{Discussion}
\label{ssec:discussion_robustLearn}

Table \ref{tab:acc_robust_learn} shows that applying LSMN (system 10) improves the ASC system accuracy increases significantly over baseline system (system 1) for both seen and unseen devices. However, for the Multi-City dataset, LSMN helps increase the accuracy on unseen cities at the cost of degraded performance towards seen cities. This indicates that there exist some acoustic scene information being stationary. The information is mixed with the location-dependent channel information. 

The HPSS method with the harmonic part discarded (system 11-13), i.e., using the percussive-enhanced log-mel feature $S_P$ as input, helps increase the accuracy on unseen cities while maintaining the performance on seen cities. The reason could be that via enhancing percussive sounds in log-mel features, the location-dependent long-duration sounds are suppressed. However, the method does not show performance gain in the Multi-Device dataset. It means that the method cannot remove the channel effect. 

The RFL framework, if applied on log-mel features decomposed by the HPSS method, does not lead to an improved ASC accuracy in general (system 14-22). As mentioned earlier, the HPSS method is a soft separation method. $S_P$ consists of both long-duration sounds and percussive sounds. The percussive sounds are suppressed instead of being totally removed. The percussive sounds do not carry device-dependent and city-dependent information, and should not be down-weighted. Down-weighting CNN learning on these percussive sounds makes the model under-fitting.


The SDBD method with long-duration component discarded (system 23-25), i.e., using $S_{short}$ as the only input, helps improve the accuracy for unseen device scenario, provided that the median filter size is $101$ or larger. By comparing system 24-25 with system 4-5, it is found that the accuracy on seen devices, seen cities, and unseen cities are decreased. This indicates that $S_{long}$ indeed contains both the channel information and the acoustic scene information.  

It can be observed that, using the SDBD method with the RFL framework (system 26-34), the accuracy on the Multi-Device dataset improves with the filter size increasing. As stated in Section \ref{ssec:discussion_tfdecomp}, this trend is related to better separation of transient sounds and long-duration sounds.

Systems with the defocus loss (system 26-28) show varying performance with different SDBD filter sizes. The ASC performance towards unseen conditions is not significantly improved, and even is deteriorating in some cases. This is probably due to that training with the defocus loss emphasizes too much on poorly classified examples compared to the RCE losses. Thus, it is unable to effectively down-weight the learning of long-duration sounds.

Systems with the RCE loss show significant performance gain. The RCE losses do not perform well when the filter size is small (i.e., $31$). The reason could be that a sound longer than $0.15$s may still be a transient sound that carries little device-dependent or location-dependent information. Down-weighting CNN learning on these transient sounds leads to model under-fitting. 

The best ASC performance towards unseen recording conditions is achieved by the systems using SDBD and RFL framework with the RCE loss. Generally speaking, a large SDBD median filter size (e.g., $201$) is suggested. A significant improvement of accuracy is found when CNNs are trained with strong down-weighting (using the RCE loss with large $\alpha$) on embedding feature learning with long-duration sounds.

\section{Conclusions and Future Work}
\label{sec:conclusions}

In this paper, we study the SDBD's duration boundary that separates long-duration and short-duration sounds. It is found that when the long-short boundary is larger than or equal to $0.15$ second, SDBD becomes effective in improving the ASC performance. For ASC with multiple devices, we observe that the increase of long-short boundary leads to increasing ASC accuracy. 

The HPSS method and SDBD are compared. The HPSS method results in a soft separation of long-duration and short-duration sounds while SDBD results in a hard separation. We find that SDBD outperforms the HPSS method in multiple configurations. 

The importance of down-weighting CNN's embedding feature learning on long-duration sounds is addressed. We propose an RFL framework to down-weight CNN learning on long-duration sounds. To adjust the degree of focus on poorly classified examples, we propose $2$ auxiliary loss functions (i.e., the defocus loss and the RCE loss). Experimental results show that the proposed RFL framework  significantly improves the ASC accuracy towards unseen devices and unseen cities. 

In the future, more detailed study could be carried out on the configuration of TF decomposition and the selection of auxiliary loss function for the RFL framework. Besides, there has been a trend to use an ensemble of models to achieve a high-accuracy ASC system. Investigation on how to integrate the proposed ASC systems into an ensemble system could be carried out.


%




\ifCLASSOPTIONcaptionsoff
  \newpage
\fi



\bibliographystyle{IEEEtran}
\bibliography{IEEEabrv,refs}

\end{document}